\documentclass[aip,jcp,preprint,longbibliography,floatfix]{revtex4-2}
\usepackage{graphicx}
\usepackage{amsmath, amssymb}% Include figure files
\usepackage{color}
\usepackage{ulem}
\usepackage{hyperref}
\newcommand{\articletitle}{
	Measuring line tension: thermodynamic integration during detachment of a molecular dynamics droplet
}
\def\ptl{\partial}
\def\ie{\textit{i.e.}, }
\def\eg{\textit{e.g.}, }
\def\kB{k_\mathrm{B}}
\def\glv{\gamma_\mathrm{LV}}
\def\gsl{\gamma_\mathrm{SL}}
\def\gsv{\gamma_\mathrm{SV}}
\def\gs0{\gamma_\mathrm{S0}}
\def\gl0{\gamma_\mathrm{L0}}
\def\gv0{\gamma_\mathrm{V0}}
\def\wsl{w_\mathrm{SL}}
\def\wsv{w_\mathrm{SV}}

\def\Wdrop{W_\mathrm{drop}}
\def\ssl{s_\mathrm{SL}}
\def\ssv{s_\mathrm{SV}}
\def\slv{s_\mathrm{LV}}
\def\Ac{A_\mathrm{c}}

\def\taul{\tau_{\ell}}
\def\etabot{\eta^\mathrm{bot}}
\def\etatop{\eta^\mathrm{top}}
\def\thetaeta{\theta(\eta)}
\def\eqref#1{(\ref{#1})}

\def\bm#1{\mbox{\boldmath $#1$}}
\def\zc{z_\mathrm{c}}

\def\rc{r_\mathrm{c}}
\def\drom{\mathrm{d}}
\def\mrd{\mathrm{d}}
\begin{document}
%\preprint{}
\title{
\articletitle
}

%\input{./authors.tex}
\newcommand{\osaka}{Department of Mechanical Engineering, Osaka University, 2-1 Yamadaoka, Suita 565-0871, Japan}
\newcommand{\tuswater}{Water Frontier Research Center (WaTUS),
	Research Institute for Science \& Technology,
	Tokyo University of Science,
	1-3 Kagurazaka, Shinjuku-ku, Tokyo, 162-8601, Japan}
\newcommand{\osakam}{Department of Mechanical Engineering, 
	Osaka Metropolitan University, 3-3-138 Sugimoto, Sumiyoshi, Osaka 558-8585, Japan}
\newcommand{\tohoku}{Institute of Fluid Science, Tohoku University, 2-1-1 Katahira Aoba-ku, Sendai 980-8577, Japan}
\newcommand{\brunel}{Department of Mechanical and Aerospace Engineering, Brunel University London, Uxbridge, UB8 3PH United Kingdom}
\author{Minori Shintaku}
\affiliation{\osaka}
\author{Haruki Oga}%
\email{haruki@nnfm.mech.eng.osaka-u.ac.jp}
\affiliation{\osaka}
\author{Hiroki Kusudo}
\email{kusudo@tohoku.ac.jp}
\affiliation{\tohoku}
%\affiliation{\osaka}
%
\author{Edward R. Smith}%
\email{Edward.Smith@brunel.ac.uk}
\affiliation{\brunel}
\author{Takeshi Omori}%
\email{t.omori@omu.ac.jp}
\affiliation{\osakam}
\author{Yasutaka Yamaguchi}
\email{yamaguchi@mech.eng.osaka-u.ac.jp}
%\homepage{http://www-nnfm.mech.eng.osaka-u.ac.jp/~yamaguchi/}
\affiliation{\osaka}
\affiliation{\tuswater}
\date{\today}

\begin{abstract}
%\add{
The contact line (CL) is where solid, liquid and vapor phases meet, and Young's equation describes the macroscopic force balance of the interfacial tensions between these three phases. These interfacial tensions are related to the nanoscale stress inhomogeneity appearing around the interface, and for curved CLs, \eg a three-dimensional droplet, another force known as the line tension must be included in Young's equation. The line tension has units of force, acting parallel to the CL, and is required  to incorporate the extra stress inhomogeneity around the CL into the force balance. Considering this feature, Bey \textit{et al.} [J. Chem. Phys. \textbf{152}, 094707 (2020)] reported a mechanical approach to extract the value of line tension $\taul$ from molecular dynamics (MD) simulations.
In this study, we show a novel thermodynamics interpretation of the line tension as the free energy per CL length, and based on this interpretation, through MD simulations of a quasi-static detachment process of a quasi-two-dimensional droplet from a solid surface, we obtained the value $\taul$ as a function of the contact angle. The simulation scheme is considered to be an extension of a thermodynamic integration method, previously used to calculate the solid-liquid and solid-vapor interfacial tensions through a detachment process, extended here to the three phase system. The obtained value agreed well with the result by Bey \textit{et al.} and show the validity of thermodynamic integration at the three-phase interface.
%}
\end{abstract}

\maketitle %\maketitle must follow title, authors, abstract and \pacs
\section{Introduction}
\label{sec:intro}
Wetting plays a key role in the behavior of a liquid especially at the nanoscale where the surface-volume ratio is large. In 1805, \citet{Young1805} proposed the following equation as the force balance exerted on the contact line (CL) of a liquid in its vapor atmosphere on a flat solid surface:
\begin{equation}
	\gsl-\gsv+\glv \cos \theta = 0,
	\label{eq:Young}
\end{equation}
where $\gsl$, $\gsv$ and $\glv$ are the solid-liquid (SL), solid-vapor (SV) and liquid-vapor (LV) interfacial tensions, respectively, and $\theta$ denotes the contact angle (CA). 
Equation~\eqref{eq:Young} is called Young's equation, and the CA is used as a measure of wettability because it can be easily measured experimentally. 
Note that Young’s original work considered the mechanical balance lateral to the solid surface, 
and could not possibly be thermodynamic as it was not invented in 1805.~\cite{Gao2009} 
\par
\begin{figure}[t]
	\begin{center}
		\includegraphics[width=1.0\linewidth]{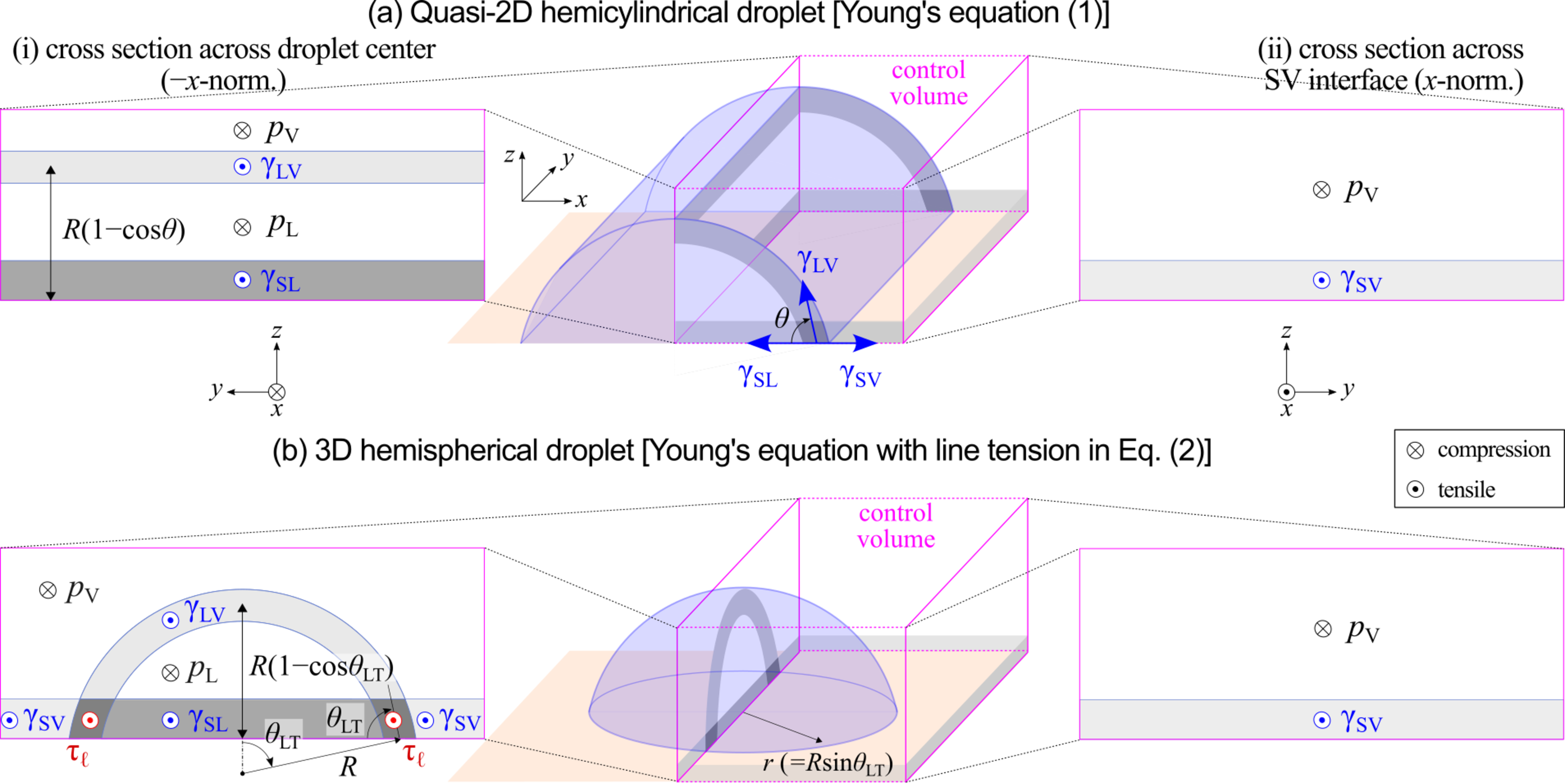}
	\end{center} 
	\caption{\label{fig:mechanical_2D3D}
		A mechanical interpretation of Young's equation \eqref{eq:Young} and the line tension modification in Eq.~\eqref{eq:Young_LT}, considering the equilibrium force balance on the rectangular control volumes (CVs) depicted in magenta.
		% (see Appendix~\ref{appsec:Young} for the details).
	}
\end{figure}

In the latter half of the 19th century, Gibbs and van der Waals formulated surface tension in the framework of thermodynamics, which is sometimes called quasi-thermodynamics because it includes the description of interfaces 
in addition to bulk.~\cite{Gibbs1961,Ono1960,Rowlinson1982,Rowlinson2002}
In this extended thermodynamic framework, Gibbs formulated surface tension as 
an excess free energy per unit area of the interface through the definition of 
the dividing surface. He also introduced the concept of line tension, 
which we express by $\taul$ here, as an excess free energy per unit length of the CL, 
\ie the force tangential to the CL.
\citet{Boruvka1977} included the effect of line tension into Young's equation
through the derivation of the variational problem of the equilibrium 
interface shape as 
\begin{equation}
	\label{eq:Young_LT}
	\gsl-\gsv+\glv \cos \theta_\mathrm{LT} + \taul \kappa= 0,
\end{equation}
where $\theta_\mathrm{LT}$ is the CA and $\kappa$ 
%in the last term LHS 
denotes the (principal) curvature of the CL, 
\eg $\kappa = 1/r$ for a 3-dimensional axi-symmetric cap-shaped 
hemispherical droplet on a flat solid surface with a circular 
CL of radius $r$ shown in Fig.~\ref{fig:mechanical_2D3D}~(b).
It follows for Eq.~\eqref{eq:Young_LT} that 
\begin{equation}
	\cos \theta_\mathrm{LT} = - \frac{\taul}{\glv}\kappa + \cos \theta,
	\label{eq:Young_LT_theta}
\end{equation}
which indicates the dependence of the contact angle 
$\theta_\mathrm{LT}$ on $\kappa$, \ie a dependence on the size of the droplet.
Unlike the surface tension $\glv$ which must be positive, 
thermodynamic arguments do not give information about the 
sign of the line tension, \ie $\taul$, 
which can be either positive or negative.~\cite{Rowlinson1982, Bey2020}
At present, 
%owing to the rapid development of high-performance computers, 
the equilibrium molecular dynamics (EMD) method can 
be used to simulate a cap-shaped hemispherical nanoscale 
liquid droplet on a solid surface. This is a powerful alternative to 
solving the variational problem in order to obtain the apparent contact angle $\theta_\mathrm{LT}$ and $\kappa$ for various sized droplets (see also Appendix~\ref{appsec:LT_sizedep}). 
Such geometrical 
analyses predicted a magnitude of $\taul$ 
which is around several pN ($\times 10^{-12}$~N),~\cite{Ingebrigtsen2007,Marchand2012}
indicating that the size effect is negligibly small 
for ordinary visible droplets.
Nevertheless, recent experimental observations of nanometer-sized 
droplets and bubbles showed that these nanodroplets and nanobubbles 
have pancake-like flat shape,~\cite{Teshima2022,Heima2023}
and it was also indicated from MD simulations that such a shape cannot be explained by simple Young's 
equation~\eqref{eq:Young}. At such scales, line tension indeed 
may play a key role.
%
%
% Considering the potential applications of nano-wetting, \eg a flow in a confined space such as a nanofoam~\cite{Rode1999} or a carbon nanotube,~\cite{Falk2014}
% the solid surfaces can have a nanoscale radius of curvature, and the interfacial tensions should depend on the curvature. 
% Regarding the water wetting on carbon nanotubes as a solid with a nanoscale curvature, unique wetting behavior~\cite{Homma2013} and a strong diameter dependence of the capillary force were experimentally reported.~\cite{Imadate2018}
%
% For the liquid-vapor interface, 
% \citet{Tolman1949} first formulated the size effect of droplet surface tension with a lengthscale called  the ``Tolman length,"~\cite{Blokhuis2006,Tumram2017,Elliott2021} and MD or MC simulations have been carried out as well.~\cite{Yaguchi2010,Das2011b,Lau2015,Cheng2018,Rehner2018,Gao2021}
%
\par
On the other hand, \citet{Kirkwood1949} developed a framework of surface tension from a viewpoint of statistical mechanics. This is a mechanical approach considering the molecular interactions based on microscopic stress description.~\cite{Irving1950} In this molecular scale, an interface is explicitly dealt with as a region with a non-zero thickness where the physical properties change continuously, and the stress is not isotropic even in static equilibrium (see also Fig.~\ref{fig:mechanical_2D3D}).
The integral of stress anisotropy around the liquid-vapor or liquid-gas interface can be related to the surface tension, a process pioneered by  Bakker.~\cite{Bakker1928, Kirkwood1949, Tolman1949, Ono1960, Rowlinson1982} Such a mechanical calculation of surface tension $\glv$ through Bakker's equation using a quasi-one-dimensional (1D) flat liquid film system 
%having two LV interface 
is considered a standard MD approach because it is easily realized by using the periodic boundary conditions (PBCs) in the surface-lateral directions.~\cite{Allen1989} 
Note that only the integral of each principal stress component in the whole system is used for the calculation of $\glv$, \ie one does not need to obtain the stress distributions which is computationally demanding and not straightforward for systems with long-range Coulomb interactions.~\cite{Shi2023}
\par
Regarding wetting including solids, beyond simple evaluation of the apparent contact angle from the shape,
a number of MD and Monte Carlo (MC) studies have been done mainly to quantitatively extract the SL and SV interfacial tensions through a thermodynamic and/or a mechanical 
approach.~\cite{Nijmeijer1990_theor, Nijmeijer1990_simul, Tang1995, Gloor2005, Ingebrigtsen2007, Das2010, Weijs2011, Seveno2013, Nishida2014, Imaizumi2020, Surblys2014, Yamaguchi2019, Kusudo2019, Leroy2009, Leroy2010, Leroy2015,  Kanduc2017, Surblys2018, Bistafa2021,  Grzelak2008, Lau2015, Kumar2014, Ardham2015, Jiang2017,  Ravipati2018, Omori2019, Watanabe2022}
Especially related to the latter, called the mechanical route, calculating the local stress distribution is one of the key issues for the understanding of wetting through the connection to macroscopic fluid mechanics.~\cite{Shi2023} 
From the visualization of the stress field in the molecular scale using a quasi-two-dimensional (2D) system achieved under the PBC,~\cite{Nishida2014, Shao2015, Kusudo2021} it has been shown that the stress anisotropy also exists at the SL and SV interfaces with finite thicknesses, and that the CL is a local region where SL, LV and SV interfaces with finite thicknesses meet and has a more complex stress features (see Fig.~\ref{fig:mechanical_2D3D} and also Fig.~\ref{fig:press} in Appendix~\ref{appsec:stress}). The present authors showed that $\gsl$ or $\gsv$ can be obtained by calculating the stress distribution along the direction normal to the solid surface away from the CL region, and proved that the expression was consistent with Young's equation~\eqref{eq:Young} by considering a control volume (CV) surrounding the CL and by determining the contact angle $\theta$ from the extrapolation of the LV interface shape.~\cite{Yamaguchi2019} 
In other words, %\add{
	the force balance on the CV faces away from the CL is considered, and
	%} 
the CL region %\add{
	having complex stress distribution
	%} 
is not explicitly included in Young's equation~\eqref{eq:Young} in this quasi-2D framework. 
\par
Related to these studies, %\add{
	in this work we provide
	%} 
one possible intuitive justification about the equilibrium force balance of Young's equation in Eq.~\eqref{eq:Young} and modified one in Eq.~\eqref{eq:Young_LT} 
%\rem{is provided in Appendix~XX%~\ref{appsec:Young}
	%,} 
%\add{
	here. %} 
%
%\add{
	Figure~\ref{fig:mechanical_2D3D} shows the CVs with one 
	face passing through the center of the quasi-2D and 3-dimensional (3D) equilibrium droplets of radius $R$ and contact angle $\theta$ or $\theta_\mathrm{LT}$, respectively. Both have a LV-interface with uniform curvature. 
	The anisotropic stress features mentioned above are schematized
	in these figures (see also Fig.~\ref{fig:press}), 
	and the CL is considered to be a region where three interfaces, each with a finite thickness, %\rem{cross each other} \add{ 
		meet.
		Now we think about the equilibrium force balance on the control volumes (CVs) shown in magenta in the center of Fig.~\ref{fig:mechanical_2D3D} with two parallel wall-normal faces; one set across the droplet center and the other set across the solid-vapor interface. 
		This setting is indeed similar to the explanation of the 
		Young-Laplace equation without solid by \citet{Berry1975}.
		The shear stress $\tau_{zx}$, defined as the stress in the $x$-direction on a face with outward normal in the $z$-direction, is zero on the top face because it is in the vapor bulk. 
		In addition, when the force from the solid is dealt with as an external force, \ie not included in stress, the stress integral on the bottom face is zero because no fluid molecules exist below this bottom face.~\cite{Yamaguchi2019,Schofield1982}
		%Furthermore, the $x$-direction component cancels out on the two $y$-normal faces with opposite outward directions. 
		Thus, under a condition with a flat and smooth solid wall where the external force from the solid which can be assumed to be zero, the force balance on these CVs is expressed by the $\tau_{xx}$ components on $x$-ourward-normal and $-x$-outward-normal faces displayed in Figs.~\ref{fig:mechanical_2D3D}~(i) and (ii). 
		%\par
		In the case of the quasi-2D droplet in Fig.~\ref{fig:mechanical_2D3D}~(a), by ignoring the thicknesses of the interfaces, 
		the force balance in Figs.~\ref{fig:mechanical_2D3D}~(a-i) and (a-ii) is written by
		\begin{equation}
			\glv + \gsl - p_\mathrm{L} R(1-\cos \theta) 
			=
			\gsv - p_\mathrm{V} R (1- \cos \theta)
			\label{eq:balance_2D_drop}
		\end{equation}
		where $p_\mathrm{L}$ and $p_\mathrm{V}$ denote the pressure values in liquid and vapor bulks, respectively. By inserting the Young-Laplace equation for a cylindrical interface
		\begin{equation}
			p_\mathrm{L}-p_\mathrm{V}
			= \frac{\glv}{R},
		\end{equation}
		original Young's equation~\eqref{eq:Young} is derived, which obviously does not include $\taul$, \ie the anisotropic stress on the CL.
		%
		%}
	\par
	%\add{
		On the other hand, in the case of the quasi-3D droplet in Fig.~\ref{fig:mechanical_2D3D}~(b),  
		the force balance in Figs.~\ref{fig:mechanical_2D3D}~(b-i) and (b-ii) is written by
		\begin{align}
			\nonumber
			&\glv \cdot 2R\theta_\mathrm{LT} 
			+  \gsl \cdot 2r%\cdot 2R\sin \theta_\mathrm{LT} 
			+ 2\taul
			- p_\mathrm{L}(R^{2}\theta_\mathrm{LT} - rR \cos \theta_\mathrm{LT})
			\\
			&= 
			\gsv \cdot 2r %2R\sin \theta_\mathrm{LT}
			- p_\mathrm{V}(R^{2}\theta_\mathrm{LT} - rR \cos \theta_\mathrm{LT})
		\end{align}
		where $\taul$ is assumed to be positive if it gives tensile force on the CL region.
		By inserting again the Young-Laplace equation for a spherical interface
		\begin{equation}
			p_\mathrm{L} - p_\mathrm{V} = \frac{2\glv}{R},
		\end{equation}
		and the following geometrical relation
		\begin{equation}
			\frac{1}{\kappa} = r = R\sin \theta_\mathrm{LT},
		\end{equation}
		Young's equation including line tension $\taul$ in Eq.~\eqref{eq:Young_LT} is derived.
		%}
	%where the force exerted on a rectangular control volume (CV) in Fig.~\ref{fig:mechanical_2D3D} is considered, which is
	%similar to the explanation of the Young-Laplace equation by \citet{Berry1975}.
	%
	\par
	Indeed, \citet{Bey2020} proposed a novel approach to extract $\taul$ from similar quasi-2D EMD systems as an extension of the mechanical approach, without the need for the local stress distribution, and examined the dependence of $\taul$ on the contact angle controlled by the solid-fluid interaction strength.
	They adopted a system with a quasi-2D droplet sandwiched between two parallel walls, \ie a system with two menisci, and obtained $\taul$ from the stress integral on the face normal to the CLs and geometric information obtained from the interface shape. Note that this stress integral corresponds to that on the $y$-normal face in Fig.~\ref{fig:mechanical_2D3D}~(a) which includes the contribution from line tension.
	\par
	\begin{figure}[t]
		\centering
		\includegraphics[width=\linewidth]{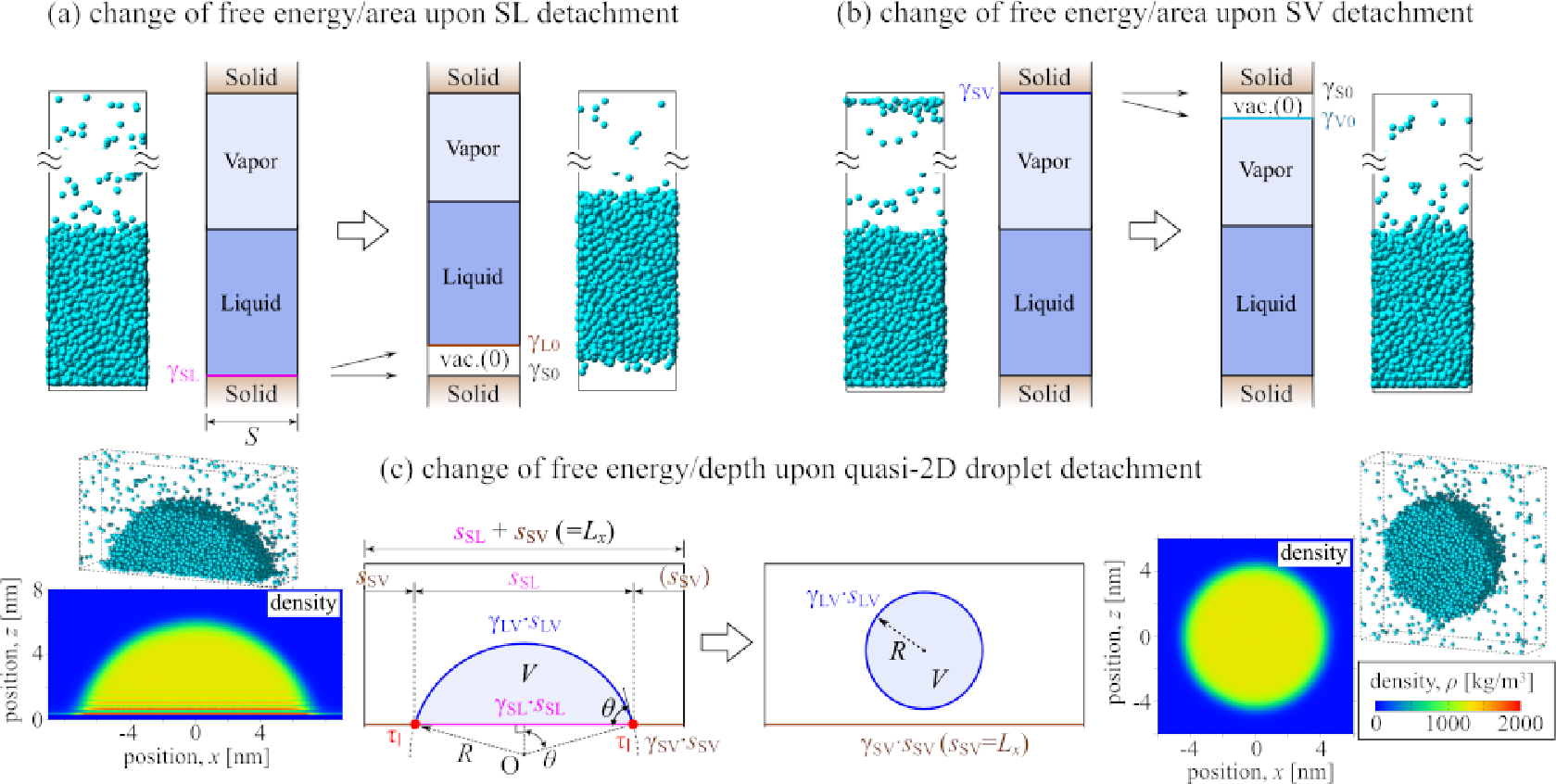}
		\caption{Schematics of the Dry-Surface (DS) method applied to quasi-1D 
			systems to calculate the (a) solid-liquid (SL), and (b) solid-vapor (SV) interfacial tensions $\gsl$ and $\gsv$, respectively from the change 
			of free energy per area obtained upon quasi-static detachment 
			of the SL and SV interfaces. 
			(c) Schematic of the DS method extended to the quasi-2D droplet 
			systems to calculate the line tension $\tau_{\ell}$.}
		\label{fig:DS-concept}
	\end{figure}
	%
	%On the other hand, 
	For the thermodynamic approach, called the thermodynamic route, the SL and SV interfacial tensions were interpreted as the interfacial free energy per %interfacial 
	area obtained through the thermodynamic integration (TI) method.~\cite{Surblys2014, Yamaguchi2019, Kusudo2019, Leroy2009, Leroy2010, Leroy2015,  Kanduc2017, Surblys2018, Bistafa2021} 
	Generally, the TI is a method to calculate the relative free energy of a target system as 
	the difference from a reference system by connecting the target and reference systems with 
	a thermodynamically reversible path using a coupling parameter embedded in the system Hamiltonian. As one possible implementation of the TI for the calculation of interfacial tension, Leroy~\textit{et al.} proposed the phantom-wall (PW) method~\cite{Leroy2009, Leroy2010} and the dry-surface (DS) method~\cite{Leroy2015} described in detail in Sec.~\ref{sec:method}. Briefly, as shown in Fig.~\ref{fig:DS-concept}~(a), a quasi-1D EMD system with a flat SL interface was used as a target system of interest in these methods, and this target system was quasi-statically substituted by a reference system with bare solid (denoted by subscript `0') and liquid surfaces along a thermodynamic path under constant number of particles $N$, temperature $T$ %and pressure $p$ ($NpT$-ensemble) or volume $V$ ($NVT$-ensemble) condition. 
	%\blue{
		and volume $V$ in a $NVT$-ensemble.
		%}
	As a result, the minimum work needed for this change 
	can be estimated as the 
	%\rem{Gibbs or} 
	Helmholtz free energy difference %\rem{$\Delta G$ or} 
	$\Delta F$ 
	%\rem{, respectively,} 
	is directly related to $\gsl$ as: 
	\begin{align}
		\nonumber
		w_\mathrm{SL}
		&\equiv \frac{\Delta F
			%\rem{(\mathrm{or\ }\Delta G)}
		}{S} 
		%  = \gs0 + \gl0 -\gsl(\eta) 
		%\\
		\label{eq:WSL_brief}
		\approx
		\gs0 + \glv -\gsl
		=
		-(\gsl-\gs0) + \glv
		,
	\end{align}
	where $\wsl$ is called the SL Work of Adhesion (WoA) as the 
	free energy per area $S$, and $\gsl - \gs0$ is the interfacial 
	free energy of SL interface relative to that of bare solid 
	surface exposed to vacuum denoted by S0.
	In the PW method, a virtual wall called the `phantom-wall' interacting only with the fluid 
	is used to strip the liquid off the solid surface by quasi-statically lifting up the phantom-wall with assigning the coupling parameter to the position of the phantom wall. This method is advantageous because it is applicable to various kinds of SL combinations with various solid-liquid interaction potential forms.~\cite{Bistafa2021, Saito2021} On the other hand in the DS method, the coupling parameter is assigned to a specific SL interaction parameter. The DS  method is powerful in the sense that the SL interfacial tension can be obtained as a semi-continuous function of the  SL interaction parameter.~\cite{Surblys2018, Yamaguchi2019}
	Similarly, the SV interfacial tension $\gsv$ can also be evaluated from the SV work of adhesion $\wsv$ using a system shown in Fig.~\ref{fig:DS-concept}~(b), which is also described in detail in Sec.~\ref{sec:method}.
	\par
	Based on these %\rem{two} \add{
		mechanical and thermodynamic
		%} 
	routes, we obtained $\gsl$ or $\gsv$ using a quasi-1D system with a flat SL or SV interface with various solid-fluid combinations, and showed that the contact angle predicted from these values corresponded well with the apparent contact angle of a quasi-2D droplet formed on the solid wall with the same solid-fluid interaction parameters.~\cite{Nishida2014,Surblys2014,Yamaguchi2019,Bistafa2021,Teshima2022}
	These studies indicated that the apparent contact angle of the droplet obtained in the MD simulations agreed well with the one predicted by Young's equation~\eqref{eq:Young} in case the solid surfaces are flat and smooth so that the CL pinning cannot be induced. 
	\par
	An important and interesting point about these results is that the contact angle of a quasi-2D droplet without including line tension $\taul$ can be estimated from the interfacial tensions $\gsl$, $\gsv$ and $\glv$ obtained by mechanical and thermodynamic approaches in quasi-1D systems without having CL. On the other hand, \citet{Bey2020} extracted line tension $\taul$ from quasi-2D systems with straight CLs of zero curvature by a mechanical approach considering the stress integral on the face normal to the CL. 
	\par
	In this study, as a thermodynamics approach, we propose %\rem{a method as} 
	an extension of the DS method to extract $\taul$ by evaluating the free energy difference from a reference system as illustrated in Fig.~\ref{fig:DS-concept}~(c) through the quasi-static 
	%\rem{adhesion} add{
		detachment
		%} 
	of a quasi-2D hemi-cylindrical droplet.
	The key concept is that we calculate the free energy difference $\Delta F_\text{drop}/L$ per unit depth ($L$: system depth) given by 
	\begin{align}
		%\begin{split}
		\frac{\Delta F_\text{drop}}{L} 
		= \Delta \left[\glv\cdot \slv\right]
		+ \Delta \left[\gsl \cdot \ssl\right]
		+ \Delta \left[\gsv \cdot \ssv\right] 
		+ 2%\rem{\Delta}
		\taul
	\end{align}
	where the free energy is given as the sum of the energy 
	of two lines $2\taul$ and interfacial enegies of LV, SL and SV interfaces with lengths $\slv$, $\ssl$ and $\ssv$, respectively.
	Note again that $\taul$ has a unit of energy/length.
	In addition, the dependence of $\taul$ on the solid-fluid interaction strength was examined, and was also compared with the result by \citet{Bey2020} and that estimated from the size dependence of the contact angle of 3D droplets.
\section{Method}
\label{sec:method}
\subsection{MD Simulation Systems}
\begin{figure}[t]
	\begin{center}
		\includegraphics[width=1.0\linewidth]{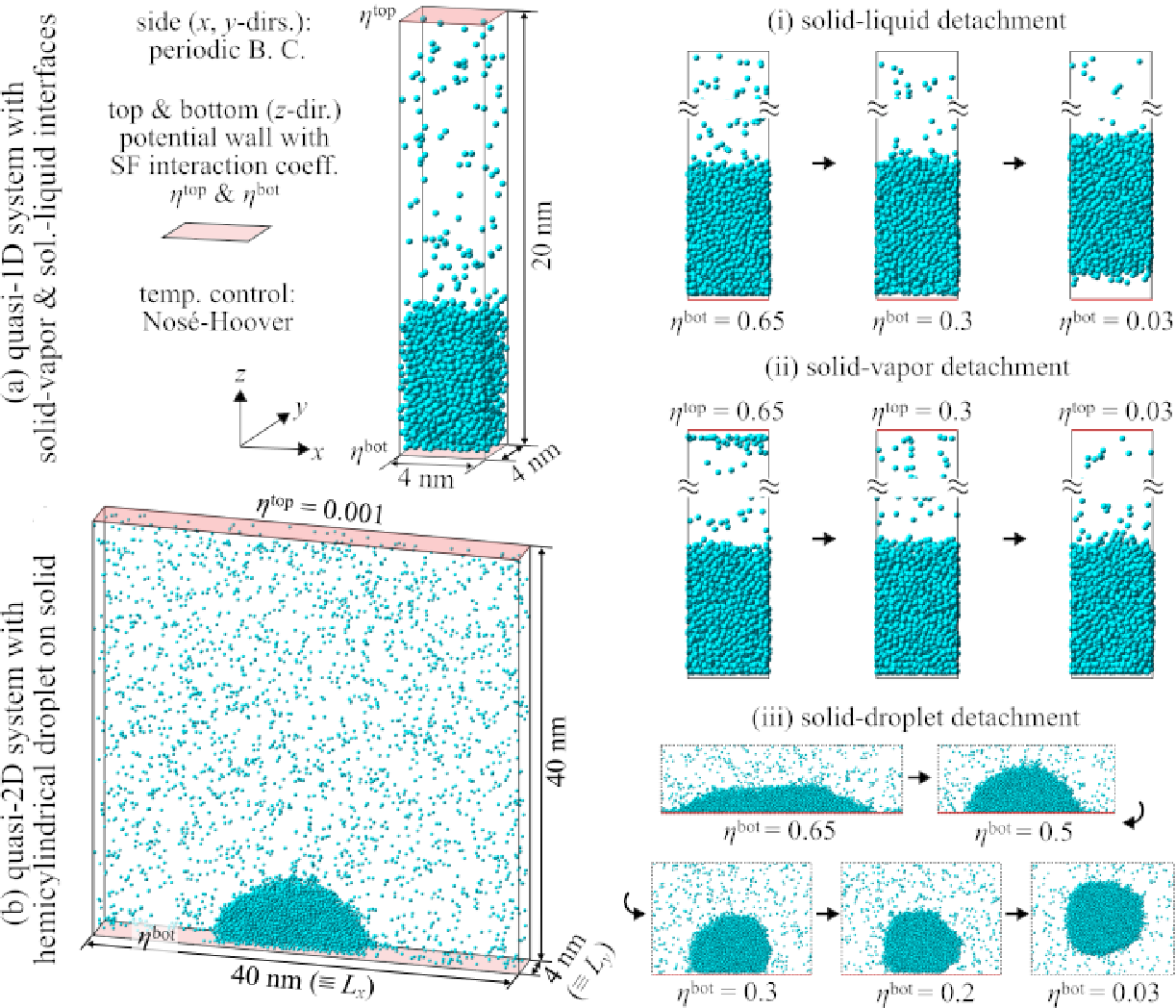}
	\end{center} 
	\caption{\label{fig:system}
		(a) Quasi-one-dimensional (1D) system with liquid and vapor attached 
		on the bottom and top solid surfaces, respectively, and schematics of 
		the (i) solid-liquid detachment, 
		and (ii) solid-vapor 
		detachment processes performed by changing the solid-fluid interaction parameter $\etabot$ or $\etatop$ of the bottom or top surface, respectively. % with keeping the other constant in this quasi-1D system. 
		(b) Quasi-two-dimensional (2D) system with a hemi-cylindrical liquid 
		droplet on the solid surface on the bottom, and (iii) schematic 
		of the solid-droplet 
		detachment process in the quasi-2D system.
	}
\end{figure}
In this study, we employed two types of equilibrium MD simulation systems: 
(a) quasi-one-dimensional systems with flat solid-liquid (SL) and solid-vapor (SV) interfaces, and 
(b) quasi-two-dimensional droplet systems with a hemi-cylindrical 
droplet on a solid surface as shown in Fig.~\ref{fig:system}. 
As the constituent fluid molecules, generic particles with the inter-particle interaction described by the 12-6 LJ potential were used. The 12-6 LJ potential expressed by
\begin{equation}
	% \nonumber
	\Phi_\mathrm{ff}(r_{ij}) \equiv
	\Phi^\mathrm{LJ}(r_{ij}) = 
	%   \Theta(
	%   \rc
	%   - r_{ij}) 
	%   \\&\times& 
	4\varepsilon_\mathrm{ff} \left[
	\left(\frac{\sigma_\mathrm{ff}}{r_{ij}}\right)^{12} 
	-
	\left(\frac{\sigma_\mathrm{ff}}{r_{ij}}\right)^{6} 
	+
	c_{2}^\mathrm{LJ}\left(\frac{r_{ij}}{\rc}\right)^2 
	+
	c_{0}^\mathrm{LJ}
	\right],
	\label{eq:LJ}
\end{equation}
was adopted for the interaction between fluid particles as a function of the distance $r_{ij}$ 
between the particle $i$ at position $\bm{r}_{i}$ and $j$ at $\bm{r}_{j}$, with $\varepsilon$ 
and $\sigma$ being the LJ energy and length parameters, respectively. A cut-off distance of $\rc=3.5\sigma_\mathrm{ff}$ was used for this LJ interaction, and by adding quadratic functions
the potential and interaction force smoothly approached zero at $\rc$.
The values of the constants $c_{2}^\mathrm{LJ}$ and $c_{0}^\mathrm{LJ}$ 
as functions of $\rc$ and $\sigma_\mathrm{ff}$ are shown in our 
previous study.~\cite{Nishida2014}
The fluid particles are expressed by `f' and corresponding 
interactions are denoted by subscripts hereafter.
\par
Both systems in Figs.~\ref{fig:system} (a) and (b) 
have two solid walls on the bottom and top of the simulation cell shown in light-red.
To minimize the number of arbitrary parameters affecting the basic physics of wetting, 
the solid wall was modeled by a simple one-dimensional 
potential field interacting with the fluid particles as a function of the distance 
rather than modeling by a group of solid particles, \eg those forming a fcc crystal. 
The interaction $\Phi_\mathrm{sf}$ %$^\mathrm{1D}$ 
between the immobile top or bottom solid wall at 
$z = z_\mathrm{s}^\mathrm{top}$ or 
$z = z_\mathrm{s}^\mathrm{bot}$, respectively 
and the fluid particle 
at $z=z_{i}$ was given by
\begin{equation}
	\label{eq:wallpot}
	\Phi_\mathrm{sf}(z'_{i};\eta)=
	%  \Theta(r^\mathrm{c}_\mathrm{sf} - z{''}_i) 
	%  \cdot 
	4\pi \rho_{n} (\varepsilon_\mathrm{sf})
	\sigma_\mathrm{sf}^{2}
	\left [ 
	\frac{1}{5} \left(
	\frac{\sigma_\mathrm{sf} }{z_{i}^{\prime}}
	\right)^{10}
	\!\!\!\! - 
	\frac{1}{2} \left(
	\frac{\sigma_\mathrm{sf} }{z_{i}^{\prime}}
	\right)^{4}
	+
	c_{2}^\mathrm{sf}
	\left(\frac{z_{i}^{\prime}}{\zc}\right)^2 
	+
	c_{1}^\mathrm{sf}
	\left(\frac{z_{i}^{\prime}}{\zc}\right)
	+
	c_\mathrm{0}^\mathrm{sf}
	\right],
\end{equation}
with
\begin{equation}
	\varepsilon_\mathrm{sf}
	=
	\eta \varepsilon_\mathrm{sf}^{0},
	\quad
	z_{i}^{\prime} = z_\mathrm{s}^\mathrm{top} - z_{i}
	\quad
	\mathrm{or} \quad
	z_{i}^{\prime} = z_{i} - z_\mathrm{s}^\mathrm{bot}.
\end{equation}
This potential field corresponds to a mean potential field created 
by a single layer of uniformly distributed solid particles with 
an area number density $\rho_{n}$, which interact with the fluid 
particles through the LJ potential with the energy and length 
parameters being 
$\varepsilon_\mathrm{sf}
(=
\eta \varepsilon_\mathrm{sf}^{0})$ 
and 
$\sigma_\mathrm{sf}$, respectively, 
where the solid-fluid (SF) interaction parameter $\eta$ was multiplied to the 
base value of $\varepsilon^{0}_\mathrm{sf}$ of $1.29\times 10^{-21}$~J as described below.
Similar to Eq.~\eqref{eq:LJ}, this potential field 
in Eq.~\eqref{eq:wallpot} was 
truncated at a cut-off distance of 
$\zc=3.5 \sigma_\mathrm{sf}$ with a quadratic function with which the potential and interaction 
force smoothly vanished at $\zc$. 
\par
The quasi-1D system in Fig.~\ref{fig:system}~(a) 
contained 2000 fluid particles in a simulation cell of $4\times 4 \times 20$~nm$^{3}$ with the top and bottom walls modeled by Eq.~\eqref{eq:wallpot} 
with $z_\mathrm{s}^\mathrm{top}=20$~nm and 
$z_\mathrm{s}^\mathrm{bot}=0$, respectively. 
On the other hand, the quasi-2D system in Fig.~\ref{fig:system}~(b) contained 7000 fluid particles in a simulation cell of 
$40\times 4 \times 40$~nm$^{3}$
with the top and bottom walls at 
$z_\mathrm{s}^\mathrm{top}=40$~nm and 
$z_\mathrm{s}^\mathrm{bot}=0$.
The periodic boundary condition was applied 
in the wall lateral $x$- and $y$-directions 
for both systems. 
\par
The system temperature was maintained at a 
constant temperature $T$ by using the 
Nos\'{e}-Hoover thermostat with an effective mass $Q$ of 
%\red{
	$3N_\mathrm{f}\kB T \tau^{2}$ with $\tau=1.0\times 10^{-12}$~s %\rem{/m$^{-1}$}%} 
applied to all fluid particles, 
where $N_\mathrm{f}$ is the number of fluid particles. 
Note that the choice of 
$Q$ had negligible effect on the results for the 
present equilibrium systems after sufficient relaxation run.
%
%   taut=1.0d-12  !should find best order by try&error
%   heatm = dble(3*nmol)*bk*tset*tauT*tauT
%   heatp=0.0d0
%   heatq=1.0d0
%   zeta=heatp/heatm
%   zetadot=3.0d0*dble(nmol-1)*bk*(temp-tset)
%
For the temperature, we have chosen $T=100$~K, 
%We tested three temperatures of $T=$ \red{100}~K, 
which is between 
the triple point and 
critical temperatures.~\cite{Mastny2007}
Note that the temperature control had no effects 
on the results since in this study  we deal with fully-relaxed 
equilibrium systems including 
the detachment processes in 
Figs.~\ref{fig:system}~(i)-(iii) explained below. 
%
%\par
The velocity Verlet method was applied for the integration of the Newtonian equation of motion with a time step of 5~fs for all systems. The simulation parameters are summarized in Table~\ref{tab:table1} with the corresponding non-dimensional ones, which are normalized by the 
corresponding standard values based on $\varepsilon_\mathrm{ff}$,
$\sigma_\mathrm{ff}$ and mass $m_\mathrm{f}$. 
%Note that the values for another system we employed in Fig.~\ref{fig:SLpin} are also shown, whose details are described in Sec.~\ref{subsec:pinninglimit}.
%
\par
The SF interaction parameter $\eta$ for the top and bottom 
walls were set at $\etatop$ and $\etabot$, 
respectively, and they were changed in a parametric 
manner except for the top wall in Fig.~\ref{fig:system}~(b) 
fixed at $\eta^\mathrm{top}=0.001$.
With the present setup, a hemi-cylindrical droplet 
was formed on the bottom wall as an equilibrium state
in the quasi-2D system in Fig.~\ref{fig:system}~(b),
and as indicated in Fig.~\ref{fig:system}~(iii),
the contact angle $\theta$ of the droplet had a 
one-to-one correspondence with the value of $\eta$ 
for the bottom wall at a given temperature, \ie
$\eta$ expresses the wettability.
This is the case for the present solid 
modeled by a potential field exerting no 
wall-tangential force on the fluid as an 
ideally smooth solid surface without 
inducing pinning of the contact line.~\cite{Nishida2014,Yamaguchi2019, Kusudo2019} 
Note that in the present quasi-2D systems, 
effects of the CL curvature can be 
neglected.~\cite{Boruvka1977, Marmur1997, Ingebrigtsen2007, Leroy2010, Weijs2011, Nishida2014, Yamaguchi2019, Kusudo2019} 
On the other hand, with a narrow lateral size, a 
quasi-one-dimensional liquid film attached on the bottom 
wall was formed as an equilibrium state
as in Fig.~\ref{fig:system}~(a)
by setting $\eta^\mathrm{bot}>\eta^\mathrm{top}$ 
with $\eta^\mathrm{bot}$ giving a droplet contact 
angle $\theta$ below 180 degrees.
%
%so that 
%The corresponding cosine of the contact angle $\cos \theta$ 
%is from $-0.9$ to $0.9$. The definition of the 
%contact angle is described later in Sec.~\ref{sec:resdis}.
%
%
%where the base value  was given by the Berthelot mixing rule as $\varepsilon^{0}_\mathrm{sf} $. 
%
%
%
\par
The physical properties of each equilibrium system with various 
$\eta$ values were calculated as the time average of 30 and 
50~ns for the quasi-1D and quasi-2D systems, respectively, 
both of which followed an equilibration run of more than 
10~ns.
\begin{table*}[!t]
	\caption{\label{tab:table1} 
		Simulation parameters and their corresponding non-dimensional values.
	}
	%\begin{tabular*}{100mm}{@{\extracolsep{\fill}}|c|c|c|} \hline\hline
	\begin{ruledtabular}
		\begin{tabular}{cccc}
			property  & value & unit & non-dim. value
			\\ \hline
			$\sigma_\mathrm{ff}$ & 0.340 & nm & 1
			\\
			$\varepsilon_\mathrm{ff}$ & $1.67 \times 10^{-21}$ & J & 1
			\\
			$\varepsilon^{0}_\mathrm{sf}$
			& $1.96\times 10^{-21}$ & J & 1.18
			\\
			$\rho_{n}$
			& $3.61^{2}$ & nm$^{-2}$ & $1.23^{2}$
			% \\
			% $\varepsilon^\mathrm{flr}_\mathrm{sf}$
			% & $0.176\times 10^{-21} $ & J & 0.106
			% \\
			% $\varepsilon^\mathrm{ceil}_\mathrm{sf}$
			% & $0.176\times 10^{-21} $ & J & 0.106
			% \\
			% $\varepsilon^\mathrm{sid}_\mathrm{sf}$
			% & $0.192\times 10^{-21} $ & J & 0.115
			\\
			$\varepsilon_\mathrm{sf}$ & 
			$\eta \times \varepsilon^{0}_\mathrm{sf}$
			\\
			$\eta$ &
			0.03 -- 0.65 & - & -
			\\
			$m_\mathrm{f}$ & $6.64 \times 10^{-26}$ & kg & 1
			%\\
			%$m_\mathrm{s}$ & - & - & -
			%\\
			%$m_\mathrm{s}$ & - & - & -
			\\
			$T$ & 100  & K & 0.827
			\\
			$N_\mathrm{f}$ (quasi-1D) & 2000  & - & -
			\\
			$N_\mathrm{f}$ (quasi-2D) & 7000  & - & -
		\end{tabular}
	\end{ruledtabular}
\end{table*}
%
%
%%%%%%%%%%%%%%%%%%%%%%%%%%%%
\subsection{Dry-Surface Method
	\label{sec:DS}}
%%%%%%%%%%%%%%%%%%%%%%%%%%%%
%
The thermodynamic integration (TI) is a method to determine the free energy difference of two equilibrium states by connecting them with a quasi-static 
path through a TI parameter embedded in the system Hamiltonian. 
Let $\lambda$ be the TI parameter, and let the target and reference 
systems correspond to $\lambda = 0$ and $\lambda = 1$ described 
by the system Hamiltonian $H(\Gamma,\lambda)$ in a constant $NVT$ 
system as a function of all positions and momenta $\Gamma$, \ie 
the phase space variable. Then, the difference of the Helmholtz 
free energy between the two systems writes
%
% \begin{equation}
	% F\left(N,V,T,\lambda\right)=-\kB
	% T\ln{Z\left(N,V,T,\lambda\right)}
	% \end{equation}
% %
% \begin{equation}
	%     Z\left(N,V,T,\lambda\right)=\frac{1}{N!h^{3N}}\int_{\Gamma}{\exp{\left[-\frac{H(\Gamma,\lambda)}{\kB T}\right]}d\Gamma}
	% \end{equation}
%
\begin{equation}
	\label{eq:def_deltaF}
	\Delta F
	\equiv
	[F(N,V,T;\lambda)]^{\lambda=1}_{\lambda=0}
	%F\left(N,V,T,1\right)-F\left(N,V,T,0\right)
	=\int_{0}^{1}{\left(\frac{\partial F}{\partial\lambda}\right)_{N,V,T} \mrd\lambda}.
\end{equation}
%
% as the integration along the thermodynamic reversible path of constant NVT. By using Eqs. [7] and [8], it follows for Eq. [9] that

% 	\Delta F=-k_BT\int_{0}^{1}{{\frac{1}{Z}\left(\frac{\partial Z}{\partial\lambda}\right)}_{N,V,T}d\lambda}=01∂H∂λdλ
% [10]

% where ∙ denotes the ensemble average. Equation [10] means that if the system Hamiltonian H is analytically differentiable with respect to \lambda, the integrand  ∂H/∂λ in the right-most hand side can be obtained for each microscopic system with a given \lambda as the ensemble average. Hence, by numerically integrating Eq. [10], the free energy difference \Delta F can be calculated. 
%
By using the relation between the Helmholtz free energy $F$ and 
the configurational partition function $Z$, it follows for Eq.~\eqref{eq:def_deltaF}
that
\begin{equation}
	\Delta F=-\kB T\int_{0}^{1}{{\frac{1}{Z}\left(\frac{\partial Z}{\partial\lambda}\right)}_{N,V,T}\mrd\lambda}=
	\int_{0}^{1}
	\left<\frac{\ptl H}{\ptl \lambda}\right>\mrd\lambda,
\end{equation}
where the angular brackets denote the ensemble average. 
If the system Hamiltonian $H$ is analytically differentiable 
with respect to $\lambda$, \ie 
$\frac{\ptl H}{\ptl \lambda}$ 
can be calculated  for each microscopic system, 
the integrand in the right-most hand side of 
Eq.~\eqref{eq:def_deltaF} as the ensemble average
with a given $\lambda$. 
Note that in practice, multiple equilibrium MD systems of 
$\lambda$ between 0 and 1 are prepared, and 
$\left<\frac{\ptl H}{\ptl \lambda}\right>$  
is calculated in each system as the time average instead of 
ensemble average assuming ergodicity.
A similar relation can be derived for the Gibbs free 
energy difference in constant $NpT$ 
systems.~\cite{Leroy2009, Leroy2010, Yamaguchi2019}
%
%\subsubsection{}
\par
\citet{Leroy2015} proposed the Dry-Surface (DS) 
scheme as one of the TI methods to calculate the SL
interfacial tension through the fluid stripping process from 
the solid surface by embedding the TI parameter $\lambda$ 
into the SF interaction potential.
Specifically in the present study, we include 
the TI parameter $\lambda$ into the SF interaction 
in Eq.~\eqref{eq:wallpot} expressed by the LJ potential as
\begin{equation}
	\Phi_\mathrm{sf}^\mathrm{DS}(z'_{i};\eta,\lambda)\equiv
	(1-\lambda)\Phi_\mathrm{sf}(z'_{i};\eta).
\end{equation}
Then, for a constant $NVT$ system, Eq.~\eqref{eq:def_deltaF} writes
\begin{equation}
	\label{eq:deltaF_DSLJ}
	\Delta F(\eta)
	=
	\int_{0}^{1^{-}}
	\left<\frac{\ptl H}{\ptl \lambda}\right>\mrd\lambda
	= -\int_{0}^{1^{-}} \left< \sum_{i=1}^{N_f} \Phi_\mathrm{sf}(z'_{i};\eta) \right> 
	\drom \lambda.
\end{equation}
As $\lambda$ approaches 1, the SF interaction is weakened, and the solid surface 
becomes `dry' for $\lambda$ slightly smaller than 1. 
This state is denoted by $1^{-}$ as the reference system 
because at $\lambda = 1$, the SF repulsion also 
becomes zero and the fluid particles can freely pass 
through the solid wall. 
By considering the relation
\begin{equation}
	\Phi_\mathrm{sf}(z'_{i};\eta)
	=
	\eta\Phi_\mathrm{sf}(z'_{i};\eta=1)  
\end{equation}
in Eq.~\eqref{eq:wallpot}, 
and by changing the integration variable 
in the right-most hand side of  Eq.~\eqref{eq:deltaF_DSLJ} 
from $\lambda$ to $\eta^{\prime}$ as 
\begin{equation}
	\eta^{\prime} = (1-\lambda) \eta,
	%   \quad
	%   \eta d \lambda = - d\eta{'}
\end{equation}
it follows
\begin{equation}
	\label{eq:deltaF_DSLJ_eta}
	\Delta F(\eta)
	=
	-\int_{0^{+}}^{\eta} \left< 
	\sum_{i=1}^{N}
	\Phi_\mathrm{sf}(z'_{i};\eta=1)
	\right> 
	\mrd \eta^{\prime}
	=
	-\int_{0^{+}}^{\eta} 
	\frac{1}{\eta^{\prime}}
	\left< 
	\sum_{i=1}^{N}
	\Phi_\mathrm{sf}(z'_{i};\eta^{\prime})
	\right> 
	\mrd \eta^{\prime},
\end{equation}
where $0^{+}$ denotes a value of $\eta^{\prime}$ 
slightly larger 
than zero. For instance, it can be set at $\eta=0.03$
with which the droplet is completely detached 
from the solid surface as displayed in Fig.~\ref{fig:system}~(iii). 
Equation~\eqref{eq:deltaF_DSLJ_eta} means that we 
get the system trajectory with the corresponding SF
interaction parameter $\eta = \xi$, 
whereas we calculate the ensemble average 
(substituted by the time average) of the 
total SF interaction potential energy 
$\sum_{i=1}^{N} \Phi_\mathrm{sf}(z'_{i};\xi)$
and divide it by $\xi$ as the integrand to 
numerically integrate the right-most hand side.
A remarkable advantage of the DS method is that 
Eq.~\eqref{eq:deltaF_DSLJ_eta} is an indefinite 
integral form of $\eta$, \ie
the free energy difference $\Delta F(\eta)$ from the 
reference system can be obtained as a semi-continuous 
function of $\eta \in [0^{+},\eta_\mathrm{max}]$, where 
$\eta_\mathrm{max}$ is the maximum value of $\eta$ 
to be investigated. More concretely, we calculate
multiple equilibrium systems with discrete $\eta^\prime$ 
values between $0^{+}$ and 
$\eta_\mathrm{max}$ with a sufficiently 
small increment $\mrd \eta^\prime$,
and calculate the integrand 
$\frac{1}{\eta^\prime}
\left< 
\sum_{i=1}^{N}
\Phi_\mathrm{sf}(z'_{i};\eta^\prime)
\right>$
in Eq.~\eqref{eq:deltaF_DSLJ_eta} as the time average
in each system, 
then,  
$\Delta F(\eta)$ is obtained as the integral from $0^{+}$ 
up to $\eta\  (\leq \eta_\mathrm{max})$ by numerical integration.
%because it is obtained upon the numerical integration 
%from the 
%above-mentioned 
%calculated in systems 
%with discrete $\eta$ values
%$0^{+}$ and $\eta_\mathrm{max}$. 
%We made use of this feature in this study to extract 
%line tension.
%
%
%\input{./3-resdis.tex}
%
\section{Results and Discussion}
\subsection{Works of solid-liquid and solid-vapor adhesion}
Figure~\ref{fig:DS-concept} shows the schematic of the DS method applied 
to quasi-1D systems to calculate the SL and SV interfacial tensions 
$\gsl$ and $\gsv$, respectively as the free energy per area obtained 
upon quasi-static detachment of the SL and SV interfaces. 
To calculate the SL interfacial tension, we carried out 
the DS process of SL detachment as shown in 
Fig.~\ref{fig:DS-concept}~(a), where $\eta^\mathrm{top}$ was 
kept constant at 0.001 whereas $\eta^\mathrm{bot}$ was changed 
from $0.1$ to $0.7$ ($=\eta^\mathrm{bot}_\mathrm{max}$). 
Upon this process, the original 
SL interface is separated into `dry' solid-vacuum  and 
liquid-vacuum interfaces. We denote this vacuum by `0' 
hereafter. Then, the work of SL adhesion 
$w_\mathrm{SL}$ defined as the free energy per 
surface area $S$ needed for this change is written as 
\begin{align}
	\nonumber
	w_\mathrm{SL}(\eta) 
	&\equiv \frac{\Delta F(\eta)}{S} 
	= \gs0 + \gl0 -\gsl(\eta) 
	\\
	\label{eq:WSL}
	&\approx
	\gs0 + \glv -\gsl(\eta) 
\end{align}
where subscript S0 denotes the bare solid surface 
without liquid or vapor adsorbed on it. 
Considering that the effect of the vapor density 
on the LV surface tension for the temperature range 
in this study is negligible, 
$\gl0$ was approximated by $\glv$.
The value of $\glv$ was obtained from a MD 
system with planar LV interfaces by a standard 
mechanical process in which the difference between
the normal stress components vertical and 
parallel to the interface was integrated around 
the LV interface, 
%\red{
	which resulted in  $\glv=7.47\times 10^{-3}$ N/m
	at $T=100$~K.~\cite{Surblys2014,Teshima2022}
	\par
	Similarly, the DS process of SV detachment
	as shown in Fig.~\ref{fig:DS-concept}~(b) was 
	carried out to calculate the SV interfacial tension, 
	where $\eta^\mathrm{bot}$ was 
	kept constant at 1.0 whereas 
	$\eta^\mathrm{top}$ was changed 
	from $0.1$ to $0.7$ ($=\eta^\mathrm{top}_\mathrm{max}$). 
	This detachment process separates the original SV 
	interface into S0 and V0 interfaces, and the work 
	of SV adhesion $w_\mathrm{SV}$ is expressed by
	\begin{align}
		\nonumber
		w_\mathrm{SV}(\eta)
		&\equiv \frac{\Delta F(\eta)}{S}
		= \gs0 + \gv0  - \gsv(\eta)
		\\
		\label{eq:WSV} 
		&\approx \gs0 - \gsv(\eta)
	\end{align}
	considering $\gv0 \approx 0$.
	Note that $w_\mathrm{SL}$ and $w_\mathrm{SV}$ 
	have the same dimension as the interfacial 
	tensions. As described later, $\gsl-\gsv$
	appearing in Young's equation can be calculated 
	from Eqs.~\eqref{eq:WSL} and \eqref{eq:WSV}
	by eliminating $\gs0$.
	\par
	\begin{figure}[t]
		\centering
		\includegraphics[width=0.8\linewidth]{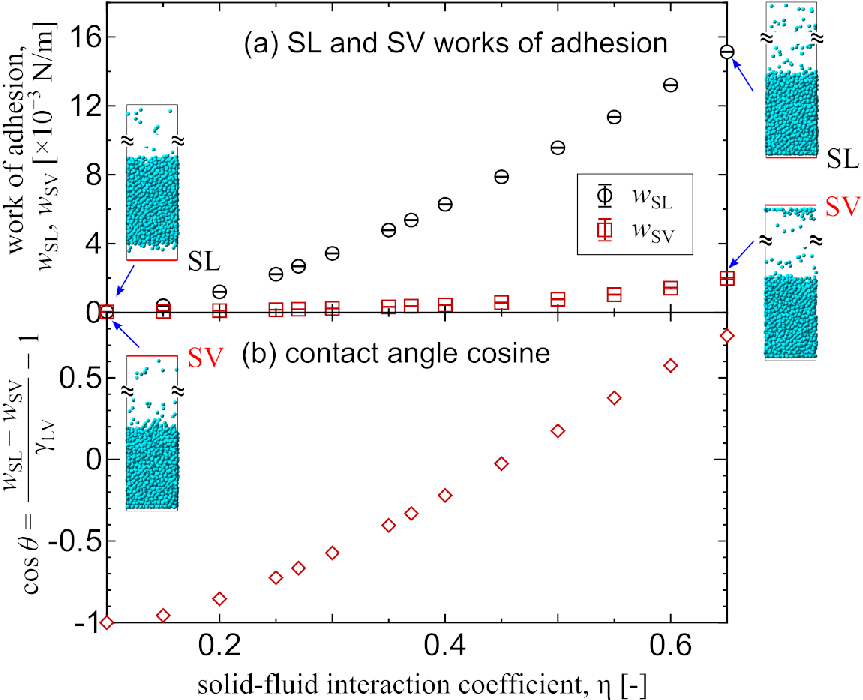}
		\caption{(a) Relation between the work of adhesion and 
			solid-fluid interaction coefficient $\eta$ 
			for SL and SV interfaces
			obtained through the DS method, and  (b) contact angle cosine estimated by the Young-Dupr\'{e} equation~\eqref{eq:Young_DS}.
			%\rem{at $T=100$~K}.
		}
		\label{fig:Wadh_SLSV}
	\end{figure}
	Figure~\ref{fig:Wadh_SLSV}~(a) shows the relation
	between the work of adhesion and 
	solid-fluid interaction coefficient $\eta$ 
	for the SL adhesion $\wsl$ and 
	the SV adhesion $\wsv$ 
	obtained through the DS method.
	With the increase of $\eta$, both works of 
	adhesion $\wsl$ and $\wsv$ became large, 
	and the work of SV adhesion had non-negligible 
	value for $\eta$ above about $0.4$, where 
	an adsorption layer was formed at the SV
	interface as observed in Fig.~\ref{fig:DS-concept}~(b).
	%\add{
		Along this quasi-static thermodynamic path, $\wsl$ and $\wsv$ were 
		obtained as smooth functions of $\eta$ through 
		the DS scheme based on Eq.~\eqref{eq:deltaF_DSLJ_eta}.
		We used these values to evaluate the contact angle $\theta$ using the Young-Dupr\'{e} equation as shown in Fig.~\ref{fig:Wadh_SLSV}~(b).
		%}
	%
	\subsection{Work of droplet adhesion}
	\par
	\begin{figure}[t]
		\centering
		\includegraphics[width=0.8\linewidth]{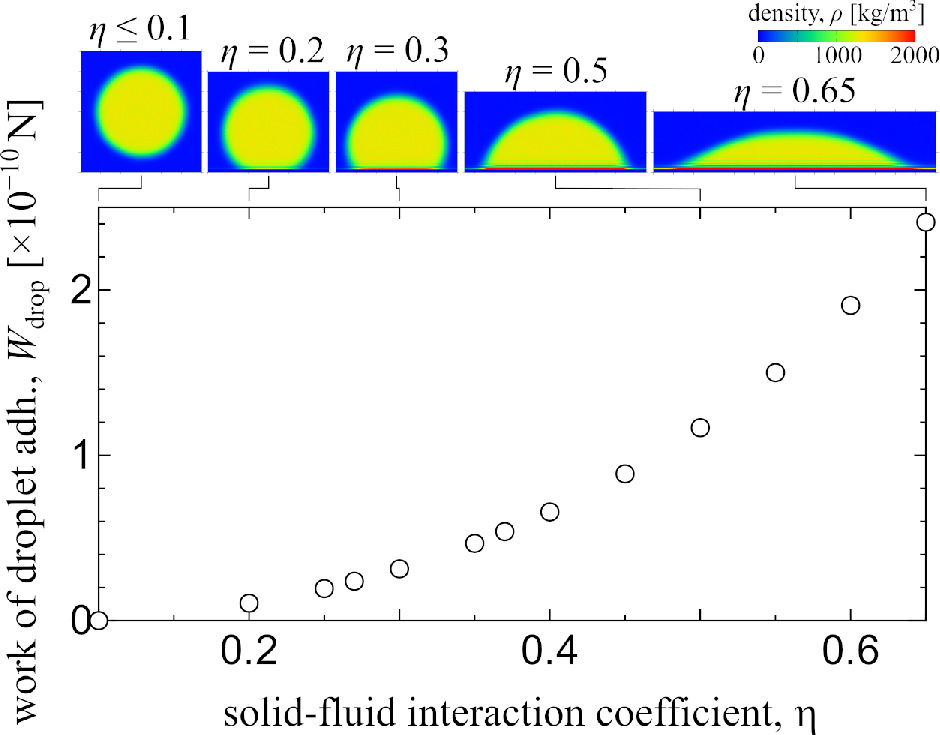}
		\caption{Work of droplet adhesion.
			% \rem{ at $T=100$~K}.
			%\add{
				Corresponding time-averaged density distributions of 
				around the center of mass of the quasi-2D droplets are shown on the top.
				The error bars are smaller than the size of symbol similar to those in Fig.~\ref{fig:Wadh_SLSV}~(a).
			}
			\label{fig:wdrop100K}
		\end{figure}
		We extended the DS method to the quasi-2D droplet 
		systems in Fig.~\ref{fig:DS-concept}~(c) to calculate 
		the line tension $\taul$. 
		Similar to the quasi-1D DS process, the droplet detachment
		process was carried out for the quasi-2D system 
		as illustrated in Fig.~\ref{fig:DS-concept}~(c), 
		where the free energy difference $\Delta F(\eta)$ 
		as the numerical integral in 
		Eq.~\eqref{eq:deltaF_DSLJ_eta} was calculated 
		using multiple equilibrium systems 
		with SF interaction $\eta$ between 
		$0^{+}$ and $\eta_\mathrm{max}$. Note that at
		$\eta=0^{+} \ll 1$, the droplet was detached 
		from the bottom wall.
		Figure~\ref{fig:wdrop100K} shows 
		the work of droplet adhesion 
		$\Wdrop$ defined as the free energy per 
		system depth $L_{y}$ needed to strip off
		the hemi-cylindrical quasi-2D droplet  
		from the solid 
		surface. Note that $\Wdrop$ has the same 
		dimension as force. Corresponding time-averaged density distributions 
		around the center of mass of the droplet for several $\eta$ values are also displayed 
		on the top panel. 
		The qualitative feature of $\Wdrop$ was the 
		same as $\wsl$ in Fig.~\ref{fig:Wadh_SLSV},
		\ie it increased with the increase of $\eta$, 
		and was obtained as a semi-smooth function 
		of $\eta$ owing to the advantage of the DS 
		method. It was also indicated from the time-averaged 
		density distributions on the top panel that 
		the LV interface away from the solid had a 
		spherical interface with a uniform curvature surface. 
		\par
		We assume that the change of bulk liquid and vapor 
		volumes upon the change of $\eta$ is negligibly small,
		\ie the total free energy of the bulk regions are 
		kept constant and the change of the system free energy 
		is due to the interface and contact line upon the 
		droplet detachment process.
		Then, $\Wdrop$ is written as
		\begin{align}
			%\begin{split}
			\nonumber
			W_\mathrm{drop} (\eta) 
			&\equiv
			\frac{\Delta F_\text{drop}(\eta)}{L_y} 
			\\
			\label{eq:Wdrop}
			&= \Delta \left[\glv\cdot \slv(\eta)\right]
			+ \Delta \left[\gsl(\eta) \cdot \ssl(\eta)\right]
			+ \Delta \left[\gsv(\eta) \cdot \ssv(\eta)\right] 
			+ 2\Delta\taul(\eta)
		\end{align}
		where $\slv$, $\ssl$ and $\ssv$ denote the lengths 
		of the corresponding interface projected in the 
		$xz$-plane, and the values with $(\eta)$ mean that 
		they depend on the SF interaction parameter $\eta$. 
		By assuming that $\glv$ is independent of $\eta$,
		\ie independent of the curvature of the LV interface 
		for the present droplet size range,~\cite{Yaguchi2010} 
		and by inserting Eqs.~\eqref{eq:WSL} and \eqref{eq:WSV}
		into Eq.~\eqref{eq:Wdrop}, it follows
		\begin{equation}
			\label{eq:Wdrop_rewrite}
			\Wdrop(\eta) 
			= \glv \cdot \Delta \slv (\eta)
			+ \wsv(\eta) \cdot L_{x} 
			- (\wsv(\eta) - \wsl(\eta) + \glv) \ssl (\eta)
			- 2\taul(\eta),
		\end{equation}
		where the simple length relation 
		$\ssl + \ssv = L_{x}$ 
		with $L_{x}$ being the system size in the $x$-direction
		is used, and we set $\taul(\eta)=0$ for $\eta \approx 0$ 
		with the droplet detached from the solid surface.
		Hence, if the two lengths $\Delta \slv$ and $\ssl$ 
		are determined as a function of $\eta$, 
		then $\taul$ can be obtained using $\wsl(\eta)$, 
		$\wsv(\eta)$ and $\Wdrop(\eta)$ as
		\begin{equation}
			\label{eq:taul=wwWll}
			\taul(\eta) = \frac{1}{2}\left[
			\glv \cdot \Delta \slv (\eta)
			+ \wsv(\eta) \cdot L_{x} 
			+ (\wsl(\eta) - \wsv(\eta) - \glv) \ssl (\eta)
			- \Wdrop (\eta) 
			\right].
		\end{equation}
		\par
		However, considering that $\taul$ is small,~\cite{Bey2020} 
		the results may depend strongly on the definition of 
		the geometric parameters
		$\Delta \slv (\eta)$ and $\ssl (\eta)$ in Eq.~\eqref{eq:taul=wwWll}.
		%
		%\add{
			To reduce the statistical error due to the fluctuation, we define %}\rem{To define} 
		these geometric parameters 
		%\rem{consistently 
			%with the works of adhesion $\wsl(\eta)$ and $\wsv(\eta)$, 
			%and also to define them as a semi-smooth function of $\eta$ 
			%as in Figs.~\ref{fig:Wadh_SLSV} and \ref{fig:wdrop100K}, 
			%we tried to} 
		%\add{
			by %}
		determining the value of droplet volume $V$, \ie the droplet area $A \equiv V/L_{y}$ 
		projected onto the $xz$-plane, 
		which we assumed to be constant independent 
		of $\eta$ in Eq.~\eqref{eq:Wdrop}
		%\add { 
			so that they are consistent 
			with the works of adhesion $\wsl(\eta)$ and $\wsv(\eta)$ and are semi-smooth functions of $\eta$ 
			as in Figs.~\ref{fig:Wadh_SLSV} and \ref{fig:wdrop100K}.
			%}. 
		As a geometrical relation (see Fig.~\ref{fig:DS-concept}~(c)), 
		the droplet area $A$ is given by
		\begin{equation}
			\label{eq:volume}
			A \equiv \frac{V}{L_{y}} =  
			R^{2} \left(
			\theta - \cos\theta\sin\theta
			\right)
		\end{equation}
		using the contact angle $\theta$ as a function of $\eta$ 
		and the radius of the 2D-droplet $R$.
		%
		%\add{
			By determining $A_\mathrm{c}$ and its error $\delta \Ac$ from the average of the projected droplet area $A$ for various $\eta$ values (see Supplementary Material), %}
		%\rem{Hence, if $A$ can be assumed to be constant at $\Ac$}, 
		the radius $R\left(\thetaeta\right)$ is determined by
		\begin{equation}
			\label{eq:R_LV}
			R\left(\thetaeta\right)=\sqrt{
				%\frac{V_\text{c}}{L_{y}}
				\frac{\Ac}{\thetaeta-\cos\thetaeta\sin\thetaeta}}
		\end{equation}
		as a function of the contact angle $\thetaeta$, and its uncertainty is also estimated using Eq.~\eqref{eq:R_LV}. 
		By using $\thetaeta$ and $R\left(\thetaeta\right)$ 
		given by Eq.~\eqref{eq:R_LV}, the 
		geometric parameters $\ssl\left(\thetaeta\right)$ and 
		$\Delta \slv\left(\thetaeta\right)$ 
		are expressed as functions of $\thetaeta$ as well by 
		\begin{equation}
			\label{eq:ssl_theta}
			\ssl\left(\thetaeta\right) = 
			2 R\left(\thetaeta\right)
			\sin\thetaeta,
		\end{equation}
		and 
		\begin{align}
			\label{eq:slv_theta}
			\slv(\thetaeta) &= 2R(\thetaeta)\cdot \thetaeta
			\\
			\label{eq:deltaslv_theta}    
			\Delta \slv(\thetaeta) 
			&\equiv 
			\slv(\pi) - 
			\slv\left(\thetaeta\right)
		\end{align}
		respectively, where $\slv(\pi)=2\sqrt{\pi \Ac}$
		is the circumference of a circle with an area 
		$\Ac$.
		Thus, if the contact angle $\thetaeta$ is obtained as 
		a semi-smooth function of $\eta$, the geomeric parameters 
		in Eqs.~\eqref{eq:ssl_theta} and \eqref{eq:deltaslv_theta} 
		can be written as semi-smooth functions of $\eta$ 
		as well, 
		and consequently, $\taul$ in Eq.~\eqref{eq:taul=wwWll} 
		is determined.
		\par
		Now, the agenda is how to obtain $\theta(\eta)$ as a 
		function of $\eta$ to be consistent with the works 
		of adhesion, and how to determine the constant volume 
		(area in the $xz$-plane) $\Ac$ in Eq.~\eqref{eq:R_LV}. 
		For the former, it has been shown in our previous 
		study that the following Young-Dupr\'{e} equation 
		holds for a droplet
		on a flat and smooth solid surface:
		\begin{align}
			\nonumber
			\cos \theta 
			&=
			\frac{-\gsl+\gsv}{\glv}
			=\frac{- (\gsl-\gs0)+(\gsv-\gs0) }{\glv}
			\\
			\label{eq:Young_DS}
			&= \frac{\wsl(\eta)-\wsv(\eta)}{\glv}
			-1,
		\end{align}
		which is rewritten by using Eqs.~\eqref{eq:WSL} and 
		\eqref{eq:WSV}. 
		The relation between $\theta$ and $\eta$ obtained 
		from Eq.~\eqref{eq:Young_DS} using 
		$\wsv(\eta)$ and $\wsl(\eta)$ in Fig.~\ref{fig:Wadh_SLSV}~(a) is shown in Fig.~\ref{fig:Wadh_SLSV}~(b).
		The contact angle indeed agreed well with the apparent contact angle, \eg estimated from the density distributions in the top panel of Fig.~\ref{fig:wdrop100K}, since the solid surface in the present study is ideally smooth.~\cite{Yamaguchi2019}
		\par
		On the other hand, for the projected area $\Ac$, 
		a difficulty exists in the definition of the 
		radius $R$ to determine the volume because 
		the interface is not a surface of discontinuity 
		but a region with a certain thickness at the 
		nanoscale.
		A possible and common choice is using the Gibbs 
		dividing surface,~\cite{Herrero2019} 
		and another choice as a strict mechanical definition 
		based on the force and momentum 
		balance was also suggested.~\cite{Yaguchi2010}
		Considering that we assume $\glv$ to be constant and also that the LV interface has a uniform curvature, 
		we used the Young-Laplace equation (quasi-2D)
		%
		% \begin{equation}
			%   \label{eq:Y-Lds}
			%   \Delta p \equiv p_\mathrm{int} - p_\mathrm{ext}
			%   =
			%   \frac{\glv}{R},
			%   \quad
			%   \therefore R = \frac{\gamma_\mathrm{LV}}{\Delta p}
			% \end{equation}
		%
		to determine the radius $R$ for each system 
		with various $\eta$%, where $p_\mathrm{int}$ and 
		%$p_\mathrm{ext}$ denote the pressure inside and 
		%outside the droplet, respectively
		(see Supplementary Material). 
		%Note that 
		%$\taul$ does not appear in the Young-Laplace equation~\eqref{eq:Y-Lds}
		%for the present hemi-cylindrical interface.
		%
		The pressure difference $\Delta p$ was estimated from the 
		stress distribution in the droplet or 
		surface normal pressure distributions on the 
		solid wall from the fluid in the present study. 
		The calculation methods and examples of the 
		stress distribution and the wall normal force 
		distribution to obtain $\Delta p$ are 
		shown in Appendix~\ref{appsec:stress}. 
		%\add{And,}
		%\add{
			%\rem{To reduce the statistical error due to the fluctuation in the wall-normal force for the present nanoscale droplets, }we estimated $A_\mathrm{c}$ and its error $\delta \Ac$ from the average of the projected droplet area $A$ for various $\eta$ values (see Supplementary Material).}
		%
		\par
		\begin{figure}[t]
			\begin{center}
				\includegraphics[width=0.8\linewidth]{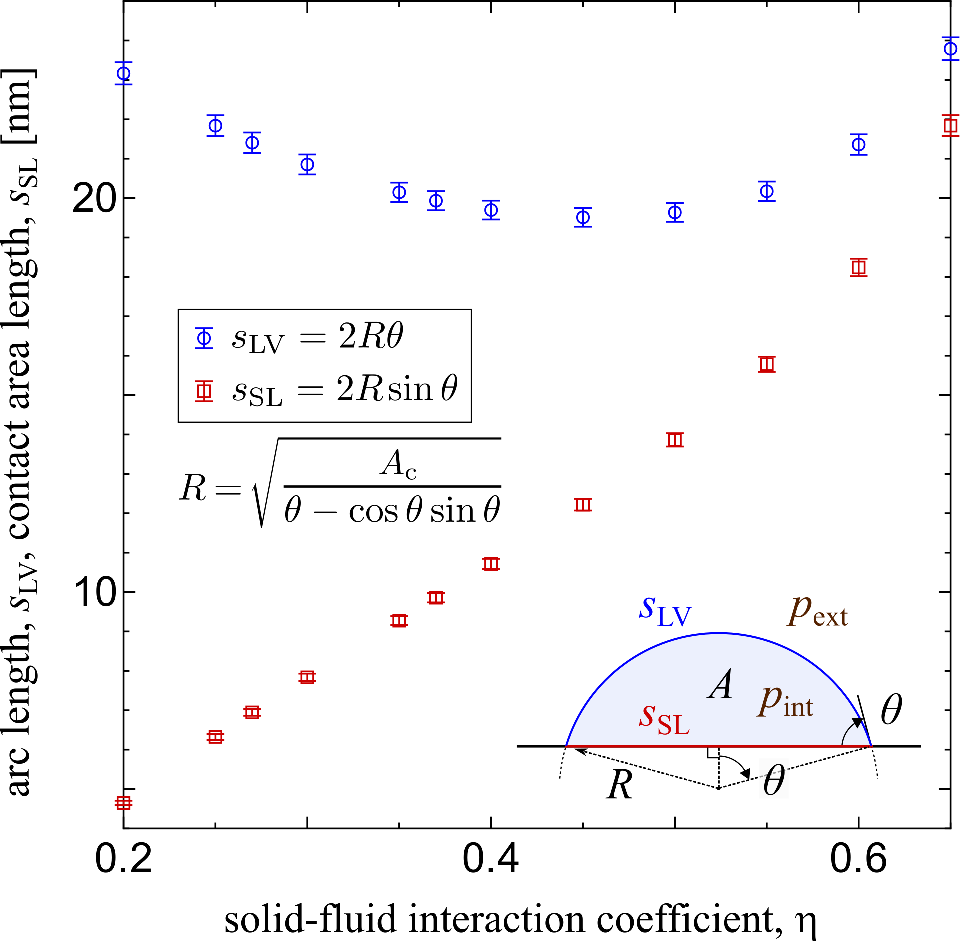}
			\end{center} 
			\caption{
				\label{fig:geometric} 
				Geometric parameters
				$\slv$ and $\ssl$ obtained each as a function of 
				$\eta$.
				% \rem{
					% (a) Relation between the solid-fluid interaction 
					% coefficient $\eta$ and cosine of the contact angle 
					% $\cos \theta$ evaluated by Eq.~\eqref{eq:Young_DS} 
					% using $\glv$ and the works of adhesion $\wsl$ and 
					% $\wsv$ in Fig.~\ref{fig:Wadh_SLSV}.
					% (b) droplet area $A\equiv V/L_{y}$ calculated by Eq.~\eqref{eq:volume}
					% using the contact angle $\theta$ and the radius $R$
					% evaluated by the Young-Laplace equation~\eqref{eq:Y-Lds}, 
					% and (c) the resulting 
					% }
				%  \rem{ The system temperature is $T=100$~K}.
			}
		\end{figure}
		\par
		Figure~\ref{fig:geometric}
		shows
		the geometric parameters $\slv$ and $\ssl$ 
		expressed as 
		functions of $\eta$. 
		As easily imagined, the contact 
		area $\ssl$ increased with the increase of $\eta$ whereas 
		the LV interface area $\slv$ showed different dependence on 
		$\eta$. For both $\ssl$ and $\slv$, the error bars mainly 
		due to the estimation of the area $\Ac$ were not 
		remarkably large.
		\par
		\begin{figure}[t]
			\centering
			\includegraphics[width=0.8\linewidth]{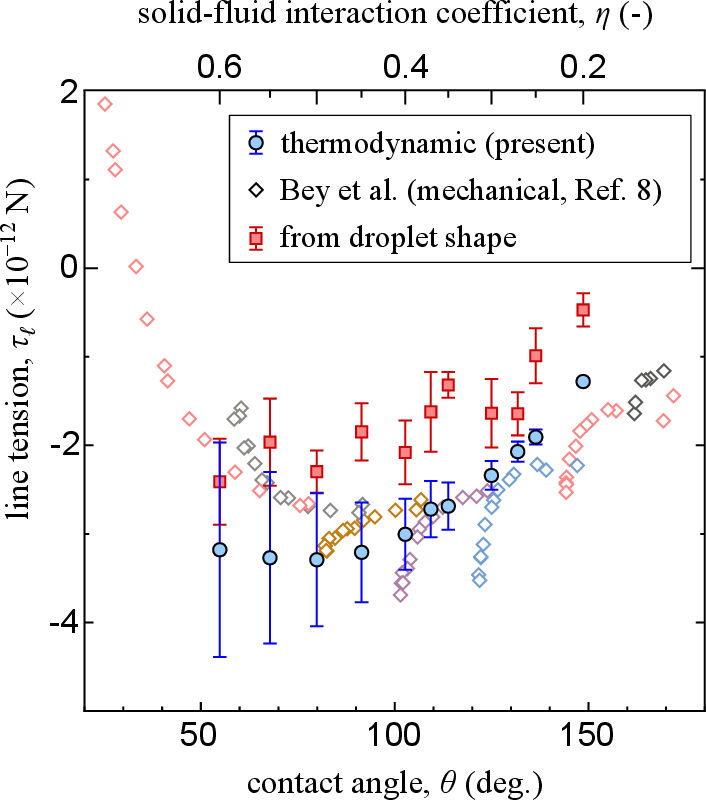}
			\caption{Line tension $\taul$ versus contact angle $\theta$
				in comparison with the result by \citet{Bey2020} 
				from a mechanical approach and that evaluated by Eq.~\eqref{eq:Young_LT_theta} from 
				the size dependence of the contact angle 
				(see Appendix~\ref{appsec:LT_sizedep})
				The corresponding $\eta$ value on the top horizontal axis is evaluated 
				by the $\theta$-$\eta$ relation in Fig.~\ref{fig:geometric}~(a).
			}
			\label{fig:tau_100K}
		\end{figure}
		%
		%Now that $\Ac$ value is determined, 
		Now, 
		all the values $R(\theta)$, $\ssl(\theta)$ and $\Delta \slv(\theta)$ in 
		Eqs.~\eqref{eq:R_LV}, \eqref{eq:ssl_theta} and \eqref{eq:deltaslv_theta}
		each as a function of the contact angle $\theta$ can be determined for a given $\theta$ value, which is directly 
		related to the solid-fluid interaction coefficient $\eta$ 
		by Eq.~\eqref{eq:Young_DS}, meaning that $\taul$ in
		Eq.~\eqref{eq:taul=wwWll} can be determined as a function 
		of $\theta$ or $\eta$.
		Figure~\ref{fig:tau_100K} shows line tension $\taul$ 
		as a function of the contact angle $\theta$
		% \rem{at $T=100$~K }
		obtained by the present thermodynamic approach, superimposed 
		on the result obtained by \citet{Bey2020} from a mechanical 
		approach and that evaluated by 
		Eq.~\eqref{eq:Young_LT_theta} from the size dependence 
		of the contact angle as a simple geometrical approach 
		(see Appendix~\ref{appsec:LT_sizedep}).
		Note that a different wall potential form was used in Ref.~\citenum{Bey2020} instead of Eq.~\eqref{eq:wallpot}.
		Corresponding $\eta$ value is displayed on 
		the top horizontal axis obtained by the relation between 
		$\eta$ and $\theta$ in Fig.~\ref{fig:geometric}~(a).
		The present results agreed well with those obtained by 
		the mechanical approach, %\blue{
			although the uncertainty in the present results is large for small contact angles.
			%}
		%for larger contact angle $\theta$, whereas the error bars of $\taul$ for the present approach became large for smaller contact angle $\theta$, \ie for large $\eta$ value, and the correspondence is not clear.
		This is because with the increase of $\eta$, both 
		$\wsl-\wsv$ and $\ssl$ become large and resulting error of 
		$(\wsl-\wsv-\glv)\ssl$ and $\Wdrop$
		in the RHS of Eq.~\eqref{eq:taul=wwWll} become large. 
		Due to this limitation, the increase of $\taul$ with the decrease
		of $\theta$ up to positive value indicated in the mechanical 
		result is not obvious.
		On the other hand, the geometric approach overall has large 
		error bars which is inevitable upon the fitting procedure 
		of $\cos \theta_\mathrm{LT}$-$\kappa$ to obtain $\taul$ 
		in Eq.~\eqref{eq:deltaslv_theta} (see Fig.~\ref{fig:CA_curv} in Appendix~\ref{appsec:LT_sizedep}).
		\par
		%\add{
			Here, we discuss about the error bars more in detail. 
			At first regarding the geometric method, it is advantageous 
			because of its simplicity, but also because of its simplicity
			it gives neither mechanical nor thermodynamic explicit insights about line tension. In addition, 
			as indicated by \citet{Ravipati2018} accurate calculation 
			of the contact angle from the density distribution may 
			need long averaging time.
			Regarding the present thermodynamic approach, in addition to 
			the problem of assuming the liquid volume to be constant 
			for wettable cases mentioned above, the error $\delta \taul$ 
			of $\taul$ depends on the error $\delta \Ac$ of $\Ac$ with
			\begin{equation}
				\frac{\delta \taul}{\delta \Ac}
				\approx
				\frac{\mrd \taul}{\mrd \Ac}
				=
				\glv\frac{\sqrt{\pi}-\sqrt{\theta-\cos\theta\sin\theta}}{2\sqrt{\Ac}},
				\label{eq:error_tau_Ac}
			\end{equation}
			and this monotonically increases with the decrease of $\theta$. Therefore, the error increase for the estimation of $\taul$ is basically inevitable %\blue{
				for small $\theta$.
				%}.  
			%, \ie with the increase of $\eta.
			It is also seen from Eq.~\eqref{eq:error_tau_Ac} 
			that the relative error $\frac{\delta \taul}{\delta \Ac}$ 
			is proportional to $\Ac^{-1/2}$, and it should decrease with 
			the increase of $\Ac$, meaning that it can be reduced by using a 
			larger system size.
			Regarding the computational cost, both the mechanical and present thermodynamic methods do not need local stress calculation which 
			is computationally demanding, and longer time averaging of the 
			ordinary equilibrium MD calculation would reduce the error for both. 
			%\rem{
				%Indeed, the present calculation time for each $\eta$ is about 
				%ten times shorter than the mechanical result,~\cite{Bey2020} 
				%and presumably, the two methods are comparable regarding the 
				%computational cost.
				%}
			%}
		%
		% \red{We set the parameter $\eta$ between 0.65 and 0.2.}
		% The error bar corresponds to the error due to the definition of the radius.
		% %
		% inevitably becomes large for large $\eta$ with small contact angle $\theta$.
		% %
		% \red{Determining the interface position is very important}
		%
		% \par
		% \begin{figure}
			%   \begin{center}
				%     \includegraphics[width=1.0\linewidth]{./fig07-tau_all.eps}
				%   \end{center} 
			%   \caption{\label{fig:tau_variousT}
				% \add{Todo(110K $\sim$ 95K)}.
				% }
			% \end{figure}
		%
		% Figure~\ref{fig:tau_variousT} shows the line tension 
		% $\taul$ for three temperatures  of $T=90$, $95$ and 
		% $100$~K.
		%
		%
%
\section{CONCLUDING REMARKS}
%\add{
In this study, we reviewed the mechanical interpretation of Young's equation where the line tension is obtained from the microscopic force balance. We then showed a thermodynamics interpretation of the line tension as the free energy per CL length, obtained from the difference between a quasi-two-dimensional hemi-cylindrical droplet on a solid surface and a cylindrical droplet with the same volume. Using this concept, we obtained the value of the line tension $\taul$ through MD simulations of a quasi-static detachment process of a quasi-2D droplet from a solid surface, an extension of the thermodynamic integration method used to calculate the SL and SV interfacial tensions individually. 
Through the comparison with the results obtained in a mechanical manner, 
it was shown that the present thermodynamic approach provided a novel way to obtain the line tension.
%}
%
\begin{acknowledgments}
%We thank Konan Imadate  for fruitful discussion. 
We cordially appreciate R. Bey, B. Coasne, and C. Picard (authors of Ref.~\citenum{Bey2020}) for prividing us their calculation data in Fig.~\ref{fig:tau_100K}. T.O., H.O., H.K. 
and Y.Y. were supported by JSPS KAKENHI grant (Nos. JP23H01346, 
JP21J20580, JP23KJ0090 and JP22H01400), Japan, respectively. 
Y.Y. was also supported by JST CREST grant (No. JPMJCR18I1), Japan.
\end{acknowledgments}
%
%\newline
\vspace{5mm} \par \noindent
\textbf{DATA AVAILABILITY}
%\newline
\par
The data that support the findings of this study are available from the corresponding author upon reasonable request.
\vspace{5mm} \par \noindent
\textbf{AUTHOR DECLARATIONS}
\newline
\textbf{Conflict of Interest}
\par
The authors have no conflicts to disclose.
\appendix
\section{Stress distribution and surface normal force on the solid wall}
\label{appsec:stress}
\begin{figure}[t]
	\begin{center}
		\includegraphics[width=\linewidth]{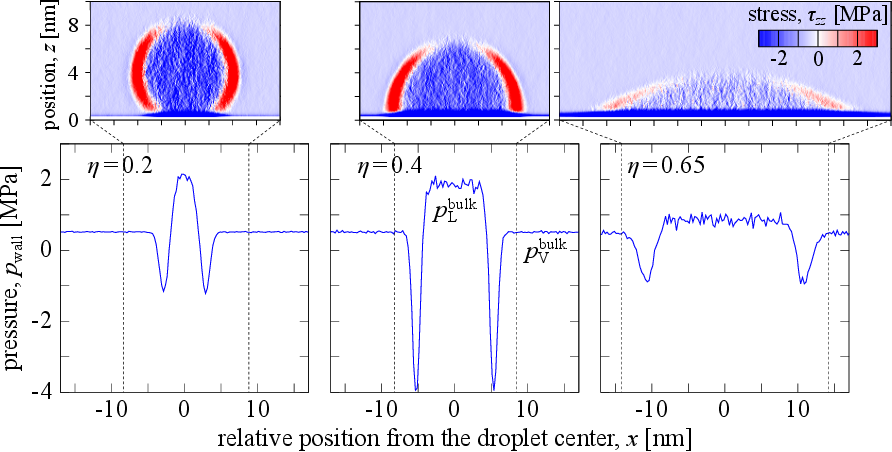}
	\end{center} 
	\caption
	{\label{fig:press} Distributions of the (top) stress component $\tau_{zz}$ in the fluid, and (bottom) pressure $p_\mathrm{wall}$ exerted on walls of $\eta=0.2$, 0.4, and 0.65 at $T=100$~K.}
\end{figure}
%
% 上記手順の1.に際して必要となる$\varDelta p$については，応力テンソルを計算して求めることが可能であるが，計算コストが高く，多くの液滴を扱う際には効率的ではない．そのため，より簡易的に求めるために，接触線近傍を除く領域において壁面が受ける圧力から求めた．図\ref{fig:press}に，$\eta=0.2,\,0.4,\,0.65$の場合のそれぞれにおける壁面が受ける圧力$p_\text{wall}$として平均時間200~nsのものを示した．図より，気液界面付近を除き，圧力の値が一定となっている領域が存在するが，これは固液・固気の界面において各々式\eqref{eq:Y-Lds}中の内圧$p_\text{int}$，外圧$p_\text{ext}$と等しい．なお図\ref{fig:press}の上部には，流体の応力テンソルの$zz$成分をそれぞれ示しているが，液相，気相バルクの応力$\tau_{zz}$(=~const.)にマイナスを掛けたものと，バルク領域における$p_\text{wall}$が対応する．壁面からはがれた円柱状の液滴系のラプラス圧に関しては，応力テンソルを計算することで求めた．
%
Figure~\ref{fig:press} shows the distributions of the stress component $\tau_{zz}$ 
in the droplet and the pressure exerted on walls of $\eta=0.2$, 0.4, and 0.65 at $T=100$~K for quasi-2D systems.
The local stress was obtained by the volume average (VA)~\cite{Nishida2014,Shi2021,Shi2023} whereas the pressure
exerted on the wall was calculated by directly time-averaging 
the local force on the wall.
Considering that the droplets showed Brownian motion on the solid surface, the distributions were taken around the center of mass of the droplet with the bin sizes of $0.02\times0.02\times4.0$~nm$^{3}$ and 0.02$\times4.0$~nm$^2$ for the VA and the wall, respectively. 
\par
The stress component $\tau_{zz}$ was homogeneous in the liquid and vapor bulk away from the interface and was inhomogeneous around the interfaces as well as around the CL. On the other hand, the pressure value $p_\mathrm{wall}$ exerted on the wall was also constant at $p_\mathrm{L}^\mathrm{bulk}>0$ around SL interface and SV interface at $p_\mathrm{V}^\mathrm{bulk}\approx 0$ interfaces while that was negative around the CL, \ie the wall was pulled upward by the LV surface tension there. The constant values $\tau_{zz}$ in the liquid and vapor bulk corresponded to the constant value of $-p_\mathrm{wall}$ at the SL and SV interfaces, respectively, and we evaluated the droplet radius $R$ from these values through the Young-Laplace equation (Supplementary Material Eq.~(S3)) with $p_\mathrm{int}=p_\mathrm{L}^\mathrm{bulk}$ 
and 
$p_\mathrm{ext}=p_\mathrm{V}^\mathrm{bulk}$ 
in this study.
\section{Line tension determined from the size dependence of 3D-droplets}
\label{appsec:LT_sizedep}
\begin{figure}
	\begin{center}
		\includegraphics[width=0.8\linewidth]{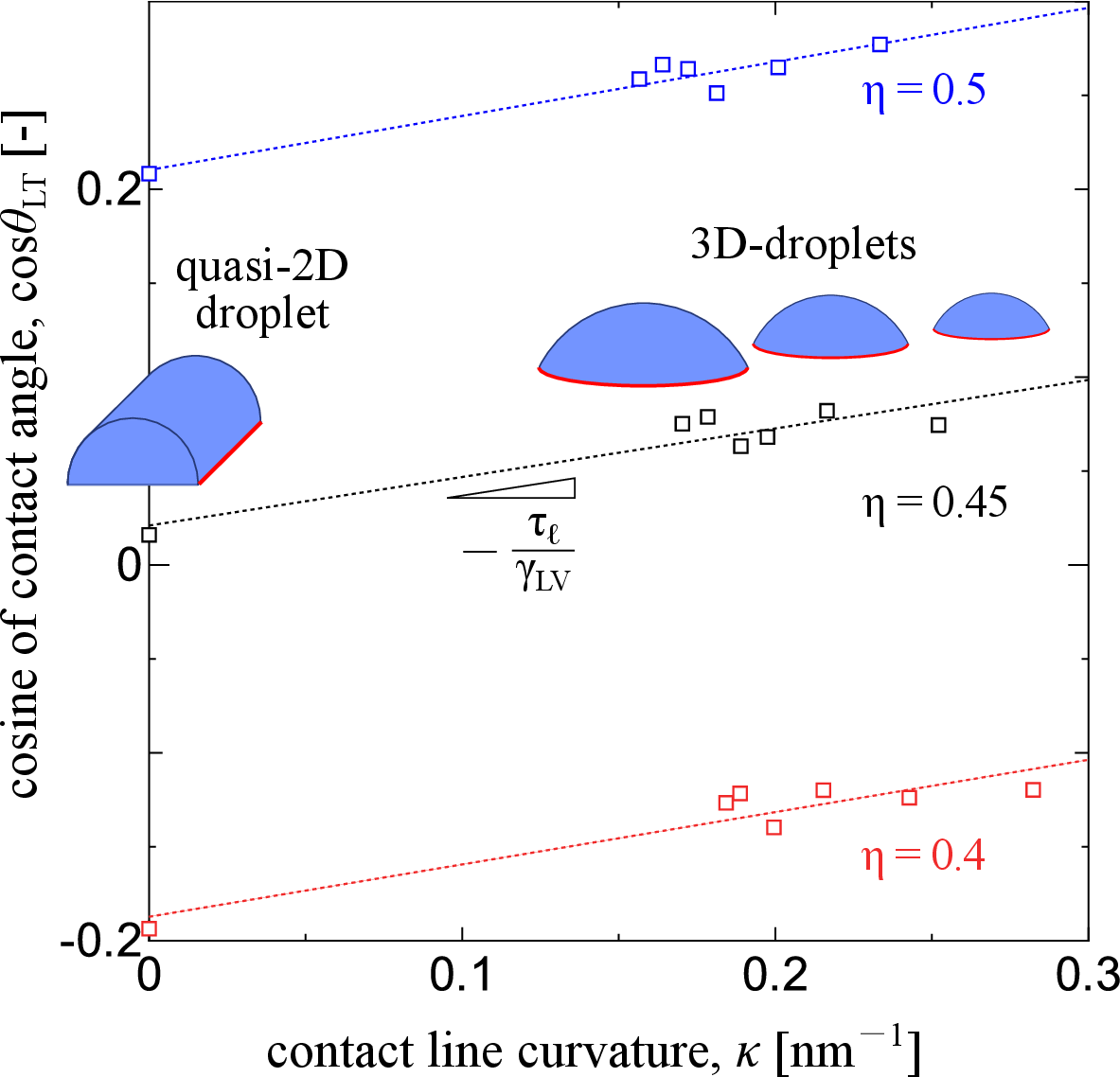}
	\end{center} 
	\caption
	{\label{fig:CA_curv} Schematic of the calculation of line tension from the size dependence of the contact angle $\theta_\mathrm{LT}$ based on Eq.~\eqref{eq:Young_LT_theta} (with the data of $\eta=0.4$, $0.45$ and $0.5$ at $T=$100~K).}
\end{figure}
Figure~\ref{fig:CA_curv} shows the schematic of the calculation of line tension from the size dependence of $\cos \theta_\mathrm{LT}$ on the CL curvature $\kappa$ based on Eq.~\eqref{eq:Young_LT_theta} with the data 
$\eta=0.4$, 0.45 and 0.5 for 3-dimensional droplets, where equilibrium droplets consisting of $N_\mathrm{f}$ fluid particles of 5,832, 6,859, 8,000, 9,261, 10,648 and 12,167 as well as for the quasi-2D droplets in the main text were used.~\cite{Ingebrigtsen2007,Marchand2012} 
We obtained the contact angles of 3-dimensional hemi-spherical (cap-shaped) droplets from the time-averaged axi-symmetric density distribution around the center of mass of the droplet.~\cite{Surblys2011} 
More concretely, the contact angle $\theta_\mathrm{LT}$ was defined as the angle between the LV interface and the solid-fluid interface plane, where $\theta_\mathrm{LT}$ was obtained by fitting a density contour at $\rho=400$~kg/m$^3$ at the LV interface away from the solid with a spherical surface with a constant curvature.~\cite{Nishida2014,Yamaguchi2019} On the other hand, the solid-fluid interface position was defined as the limit position nearest to the solid that the fluid molecule could reach.~\cite{Yamaguchi2019}
From the geometric information of the hemi-spherical droplet radius $R$ and the contact angle 
$\theta_\mathrm{LT}$, 
we evaluated the CL curvature $\kappa=1/R \mathrm{sin}\theta_\mathrm{LT}$. % line radius $r=1/\kappa$. 
By fitting the data with a straight line including the contact angle $\theta$ quasi-2D droplet as $\theta_\mathrm{LT}$ at $\kappa=0$, we obtained the values of $\taul$ for various solid-fluid interaction parameters $\eta$ in Fig.~\ref{fig:tau_100K}. 
%\add{
	As indicated from the figure, the fitting includes points of 3-dimensional hemi-spherical droplets with $\kappa$ ($=1/r$ with $r$ being the CL radius) around 0.2~nm$^{-1}$ and a point of a quasi-2-dimensional hemi-cylindrical droplet with $\kappa=0$, and the resulting uncertainty becomes inevitably large for this geometric approach.
	%}
%
%
%\bibliography{./reference.bib}

%
%
%\newpage
%\input{./SM.tex}
%
\end{document}